\documentclass[a4paper,11pt,preprintnumbers]{article}
\usepackage{pos}
\usepackage{graphicx}
\usepackage{float}
\usepackage{amsmath}
\usepackage{caption}
\usepackage{xcolor}
\usepackage{subcaption}
\usepackage{fancyhdr}

\title{Disentangling Jet Modification}

\author[a,b]{Jasmine Brewer}
\author*[a]{Quinn Brodsky}
\author[a,b]{Krishna Rajagopal}

\affiliation[a]{Massachusetts Institute of Technology,\\
  77 Massachusetts Avenue, Cambridge, USA}

\affiliation[b]{Center for Theoretical Physics, Massachusetts Institute of Technology,\\
77 Massachusetts Avenue, Cambridge, USA}

\emailAdd{jtbrewer@mit.edu}
\emailAdd{quinnb@mit.edu}
\emailAdd{krishna@mit.edu}

\abstract{Jet modification in heavy-ion collisions is an important probe of the nature and structure of the quark-gluon plasma (QGP) produced in these collisions and also encodes information about how the wakes that jets excite in a droplet of QGP form and relax.  However, in experiment, one cannot know what a particular jet seen in a heavy ion collision would have looked like without quenching, making it difficult to interpret measurements in terms of individual jet modification. The goal of this Monte Carlo study is to gain insight into the modification of jet observables using the hybrid strong/weak coupling model of jet quenching as a test bed. In this Monte Carlo study (but not in experiment) it is possible to watch {\it the same jet} as it evolves in vacuum or in QGP. We use this ability to disentangle the effects of modification of individual jets in heavy ion collisions vs.~the effects of differing selection bias on the distribution of two observables: the fractional energy loss and the groomed $\Delta R$. 
We find that in the hybrid model the distribution of groomed $\Delta R$ appears to be unmodified in a sample of jets selected after quenching, as in heavy ion collisions, and confirm that this lack of modification arises because of a selection bias toward jets that lose only a small fraction of their energy. If instead we select a sample of jets in a way that avoids this bias, and then follow these selected jets as they are quenched, we show that 
there is, in fact, a substantial modification of the $\Delta R$ of individual jets. Furthermore, we show that this jet modification is principally due to the incorporation of particles coming from the wake that the parton shower excites in the plasma as a component of what an experimentalist reconstructs as a jet. The effects we discuss, both those due to selection bias and those due to jet modification coming from the wake in the plasma, are substantial in magnitude, suggesting that our qualitative conclusions are more general than the Monte Carlo study in which we obtain them.  
}

\FullConference{%
  HardProbes2020\\
  1-6 June 2020\\
  Austin, Texas\\
  ~\\
  MIT CTP-5231}

\begin{document}
\maketitle


{\bf Introduction:} Jet quenching is an important probe of the QGP produced in heavy ion collisions, 
but quantitative comparisons are difficult because 
the modification of jets by the plasma 
necessarily introduces biases in the selection of jets.
In particular, since the production rate of jets drops rapidly as a function of 
jet energy, when one selects jets with a given energy that have propagated through QGP -- that have been quenched -- one is selecting a sample of jets biased toward those that have 
lost the least energy.  We use a Monte Carlo study to see that this also biases in favor of selecting those jets that are least modified in other respects,
and to show that those jets which lose more energy are significantly modified
in other respects too.
In the hybrid model~\cite{Casalderrey-Solana:2014bpa,
Casalderrey-Solana:2019ubu}, it is possible to compare a jet as it is before or after quenching. This provides us the opportunity to disentangle jet modification from selection bias by looking at a quenched jet and then matching
it to its unquenched version. We can either select a sample of quenched jets (as experimentalists do in heavy ion collisions) and compare them to the same jets as they would have been in the absence of any QGP, e.g. unquenched, or we can select a sample of unquenched jets (as experimentalists do in proton-proton collisions) and watch what would have happened to those same jets if they were quenched.  
Because selecting quenched jets with a given energy introduces selection bias, the comparisons we obtain via these two different procedures look very different.
Hence, drawing conclusions without first disentangling jet modification from selection bias can lead to misconceptions.



\begin{figure}[t]
\centering
  \includegraphics[scale=.3]{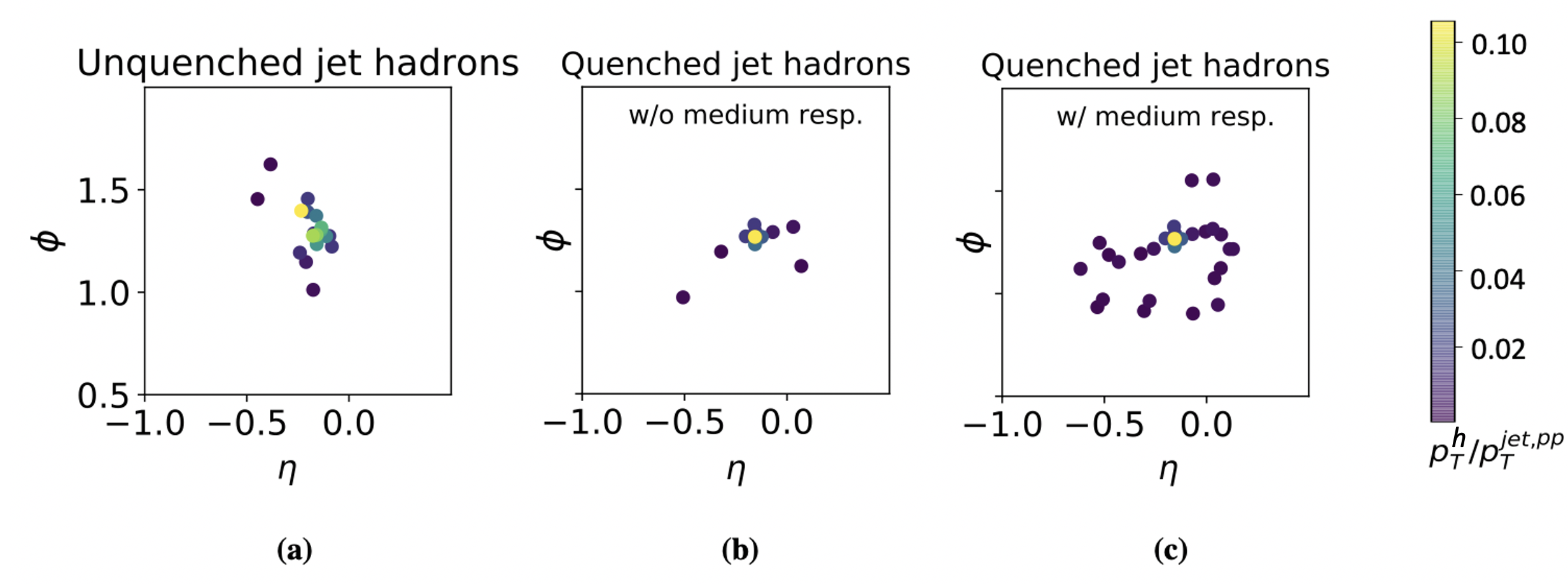}  
  \vspace{-0.06in}
\caption{(a) $(\eta,\phi)$ distribution of hadrons in an unquenched jet; (b) the same jet, after quenching, leaving out hadrons originating from the response of the medium, e.g. the wake the jet excites; (c) the same jet, after quenching, with medium response. Color indicates hadron $p_T$ as a fraction of the $p_T$ of this unquenched jet.}
\vspace{-0.09in}
\label{fig:matching}
\end{figure}

{\bf Methods:} In the hybrid model, we start with a vacuum \textsc{Pythia} shower, and embed it in an evolving boost-invariant hydrodynamic medium~\cite{Casalderrey-Solana:2014bpa}. The energy loss  of partons in the medium is determined by a formula inspired by strongly-coupled energy loss in holography~\cite{Casalderrey-Solana:2014bpa}. By turning energy loss off, one can look at the unquenched version of any individual quenched jet in the Monte Carlo ensemble. In this analysis, hadronization effects and the wake that the jet leaves behind in the medium through which it propagates were included~\cite{Casalderrey-Solana:2014bpa}. In this Monte Carlo study, we can distinguish between particles in the reconstructed jet coming from the \textsc{Pythia} shower and particles produced from medium response. We reconstruct jets using the anti-$k_T$ algorithm with $R=0.4$. 
We compare jets in the quenched and unquenched versions of the same event, matching those that are closest in the $(\eta,\phi)$ plane and within $\sqrt{\Delta \eta ^2 + \Delta \phi^2}\leq 0.4$ of each other. Fig.~\ref{fig:matching} shows an example.

In this way we can obtain a sample of the same jets before and after quenching. There are two ways to do this, which  makes it possible to study how jet selection, and jet modification each impact observables. First, we select quenched (heavy ion) jets whose transverse momentum $p_T$ lies above a threshold cut, which in our Monte Carlo study we can do with or without the particles coming from medium response included in the jets, and then for each quenched jet we use the matching procedure to find the corresponding unquenched (proton-proton) jet.  We'll call this the ``Quench-then-Select Method''.
In the alternative ``Select-then-Quench Method'', we select proton-proton jets whose $p_T$ lies above a threshold and then use the matching procedure to find the heavy ion jet that the pp jet becomes after being quenched according to the hybrid model.



\begin{figure}[t]
\begin{subfigure}{.45\textwidth}
  \includegraphics[scale=.30]{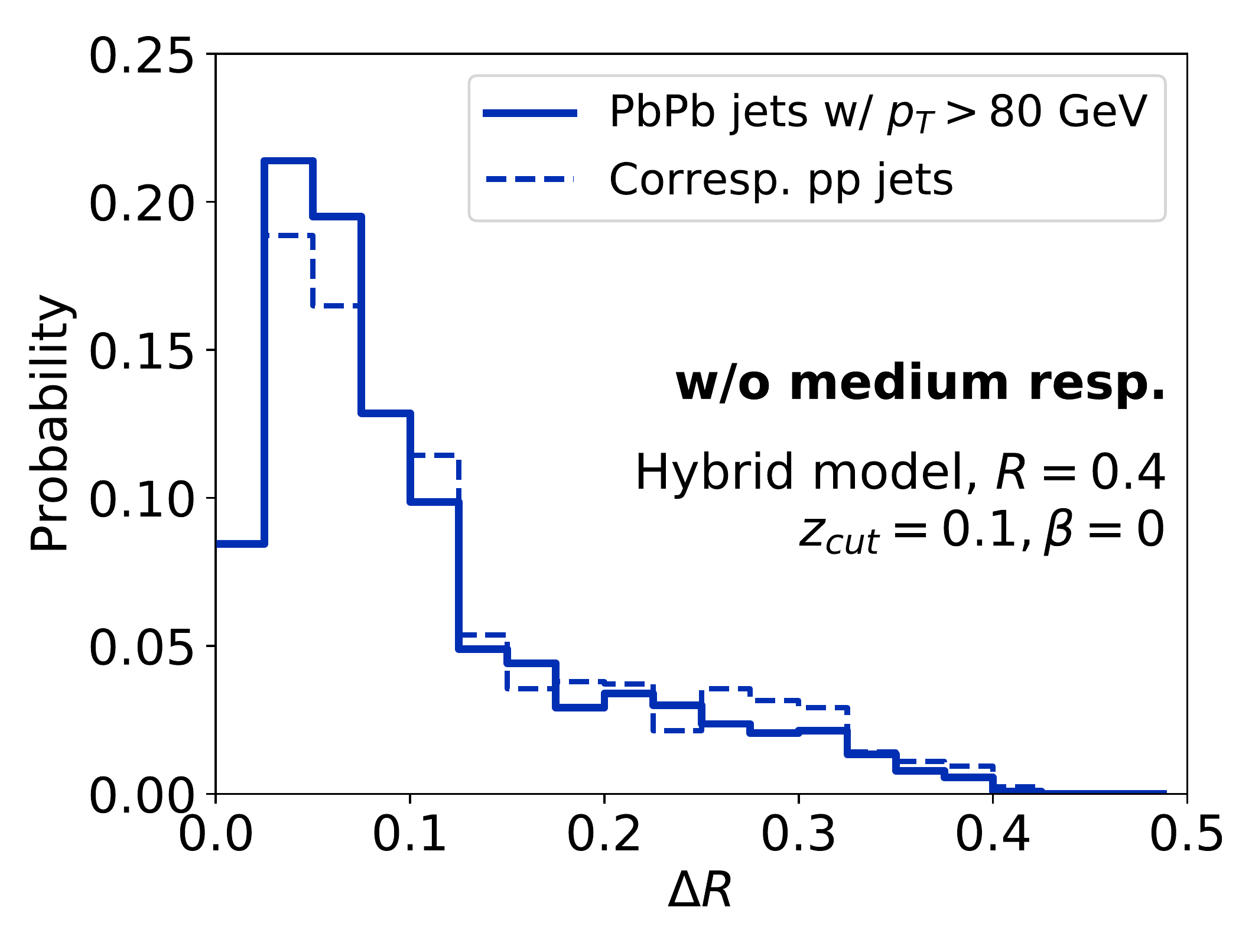}  
  \caption{}
  \label{fig:deltaR_method1_nomed}
\end{subfigure}
\begin{subfigure}{.45\textwidth}
  \includegraphics[scale=.30]{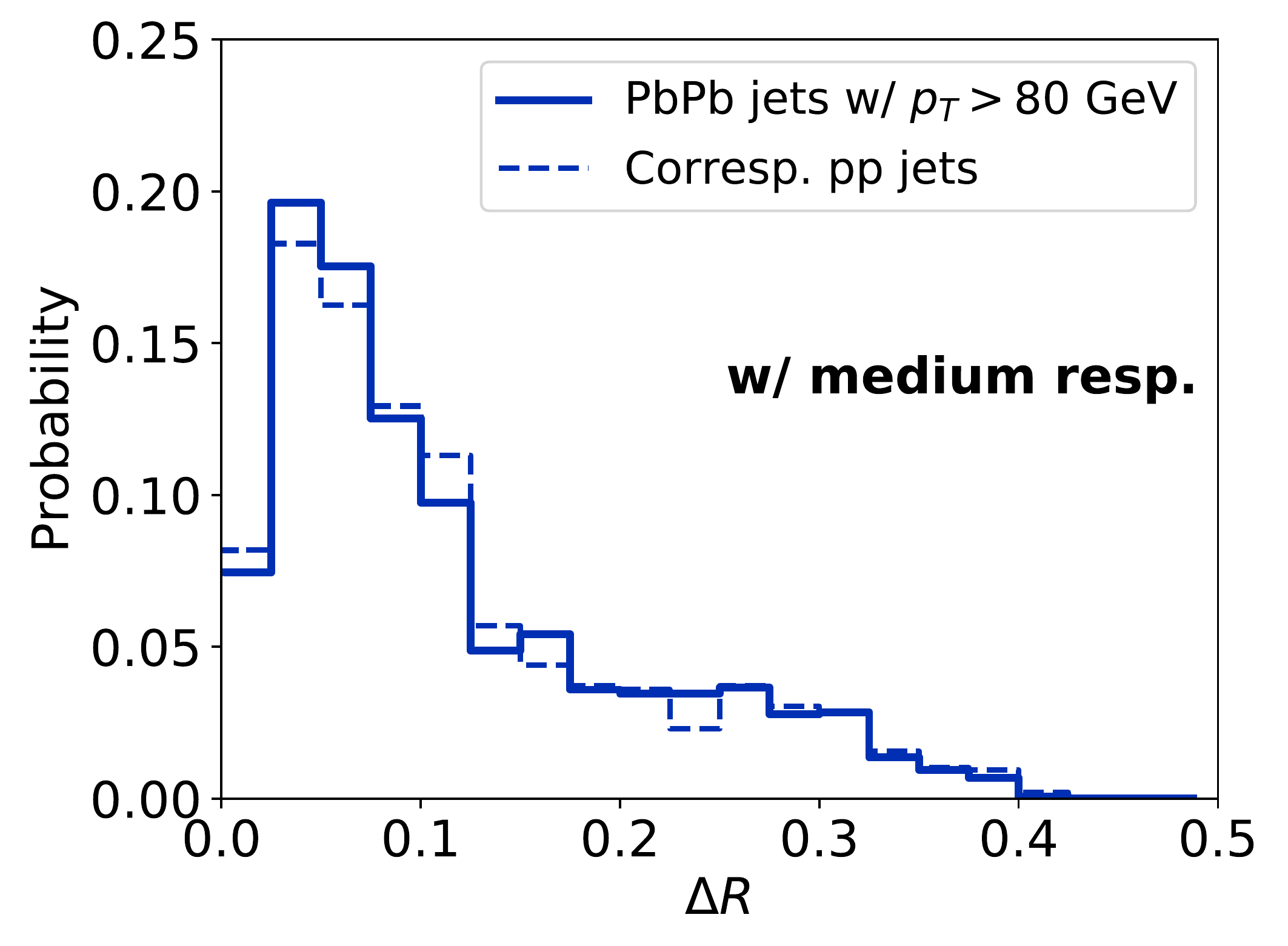}  
  \caption{}
  \label{fig:deltaR_method1_med}
\end{subfigure}
\centering
\begin{subfigure}{.45\textwidth}
  \includegraphics[scale=.30]{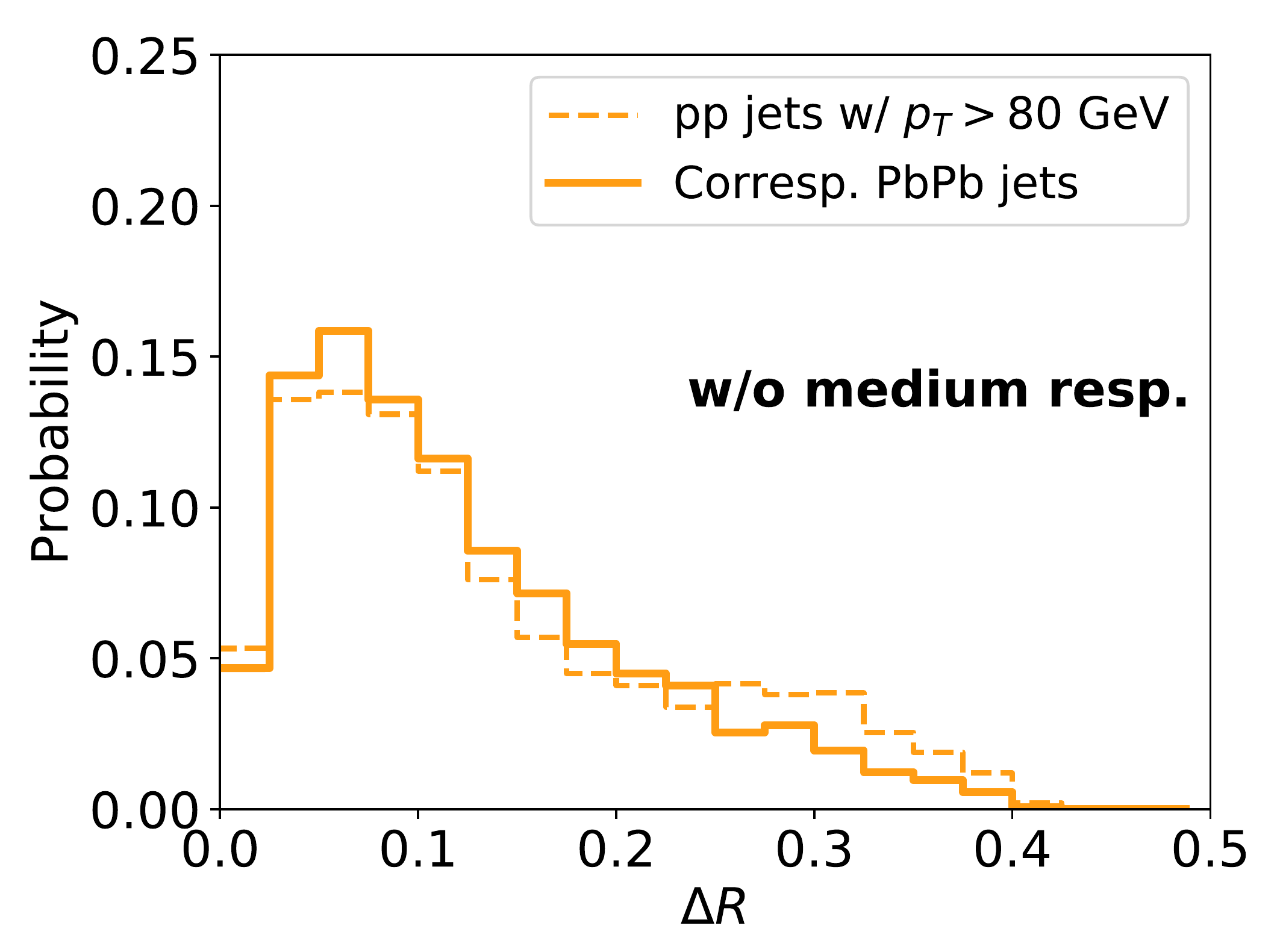}  
  \caption{}
  \label{fig:deltaR_method2_nomed}
\end{subfigure}
\begin{subfigure}{.45\textwidth}
  \includegraphics[scale=.30]{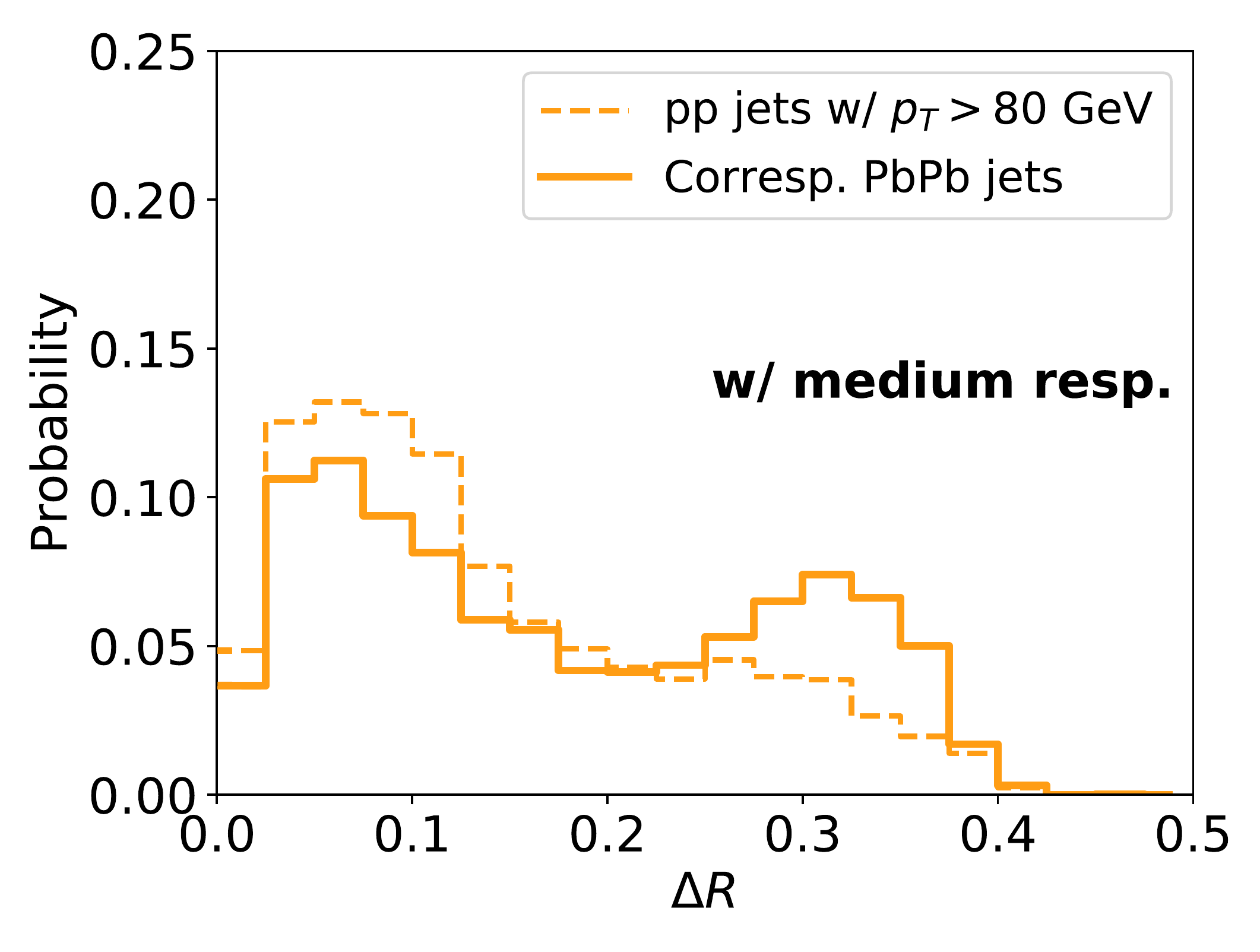}  
  \caption{}
  \label{fig:deltaR_method2_med}
\end{subfigure}
\centering
\vspace{-0.06in}
\caption{Softdrop $\Delta R$ distributions for Quench-then-Select Method (blue; upper panels) and Select-then-Quench Method  (orange; lower panels), in both cases without (left panels) and with (right panels) medium response. Upper panels  are PbPb jets without (a) and with (b) medium response,
selected with $p_T>80$ GeV (dashed) and the same jets before quenching (solid). Lower panels are pp jets selected with $p_T>80$ GeV (dashed), and the same jets  (solid), without (c) and with (d) medium response.}
\vspace{-0.09in}
\label{fig:deltaR}
\end{figure}

{\bf Results:} If the effect of the jet selection were small, one might anticipate that these methods would produce similar results since in both cases we compare quenched and unquenched versions of the same jet. However, by plotting various observables, we see that this is not at all the case. We show the Soft Drop $\Delta R$ distributions for jets with $z_{\text{cut}}=0.1$ and $\beta=0$~\cite{Larkoski:2014wba} in Fig.~\ref{fig:deltaR}, with and without medium response.
The plots using the Select-then-Quench method (Figs.~\ref{fig:deltaR_method1_nomed} and \ref{fig:deltaR_method1_med}) 
are similar to what has been seen at the partonic level 
in previous hybrid model studies of
jets selected in heavy ion collisions~\cite{Casalderrey-Solana:2019ubu}, and seem to indicate that quenching hardly modifies the $\Delta R$ distribution. The lower panels in Fig.~\ref{fig:deltaR}, though point to a completely different conclusion.


\begin{figure}[t]
\begin{subfigure}{.45\textwidth}
  \includegraphics[scale=.40]{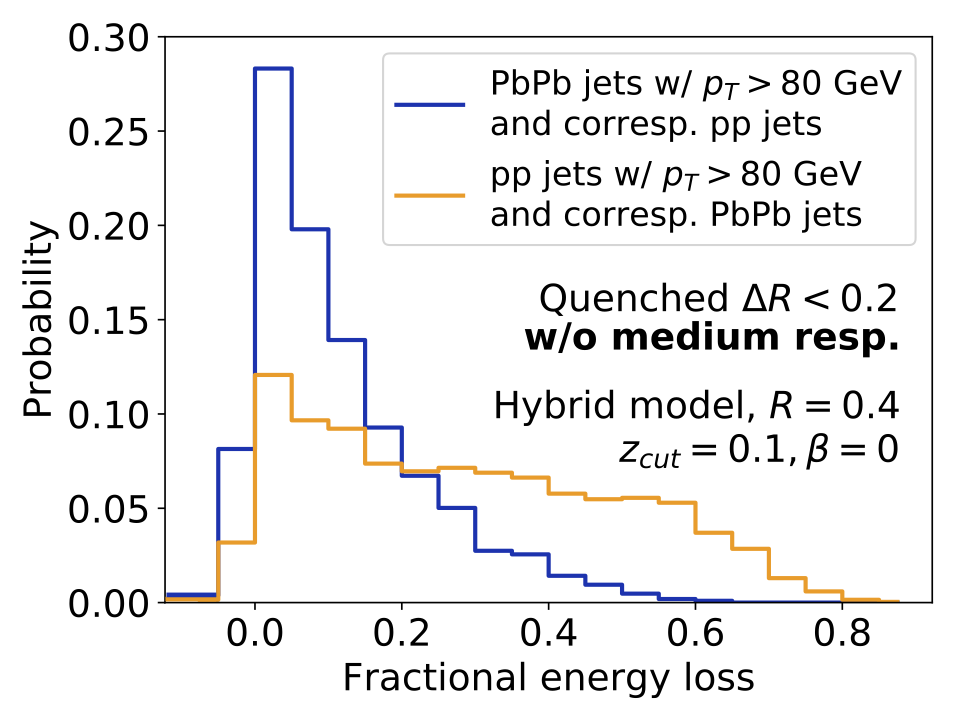}  
  \caption{}
  \label{fig:Eloss_smallR_nomed}
\end{subfigure}
\begin{subfigure}{.45\textwidth}
  \includegraphics[scale=.40]{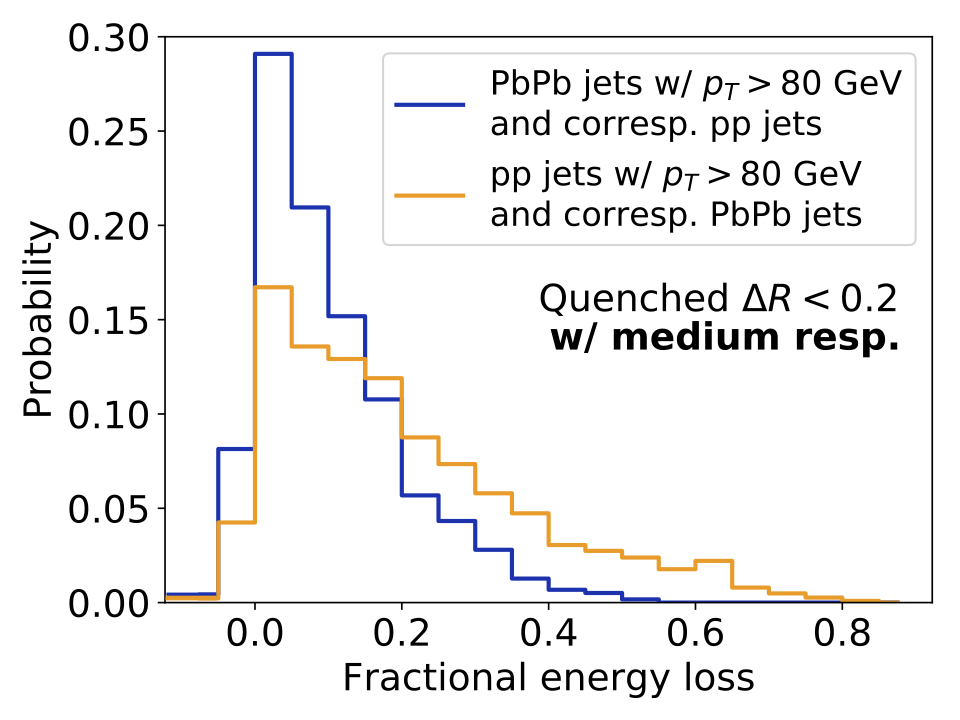}  
  \caption{}
  \label{fig:Eloss_smallR_med}
\end{subfigure}
\centering
\begin{subfigure}{.45\textwidth}
  \includegraphics[scale=.40]{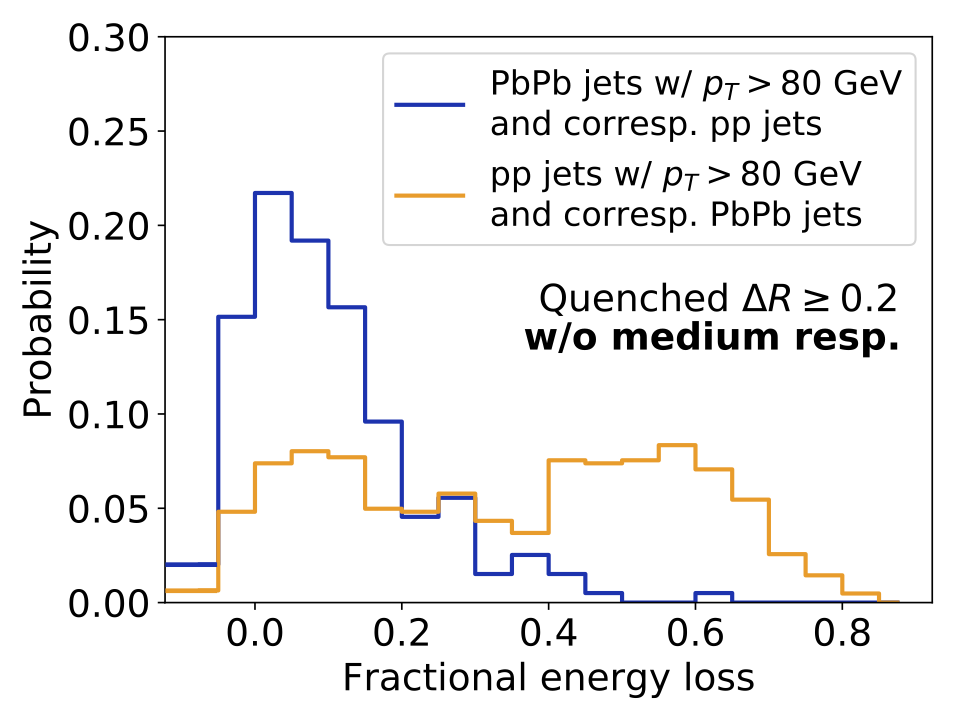}  
  \caption{}
  \label{fig:Eloss_largeR_nomed}
\end{subfigure}
\begin{subfigure}{.45\textwidth}
  \includegraphics[scale=.40]{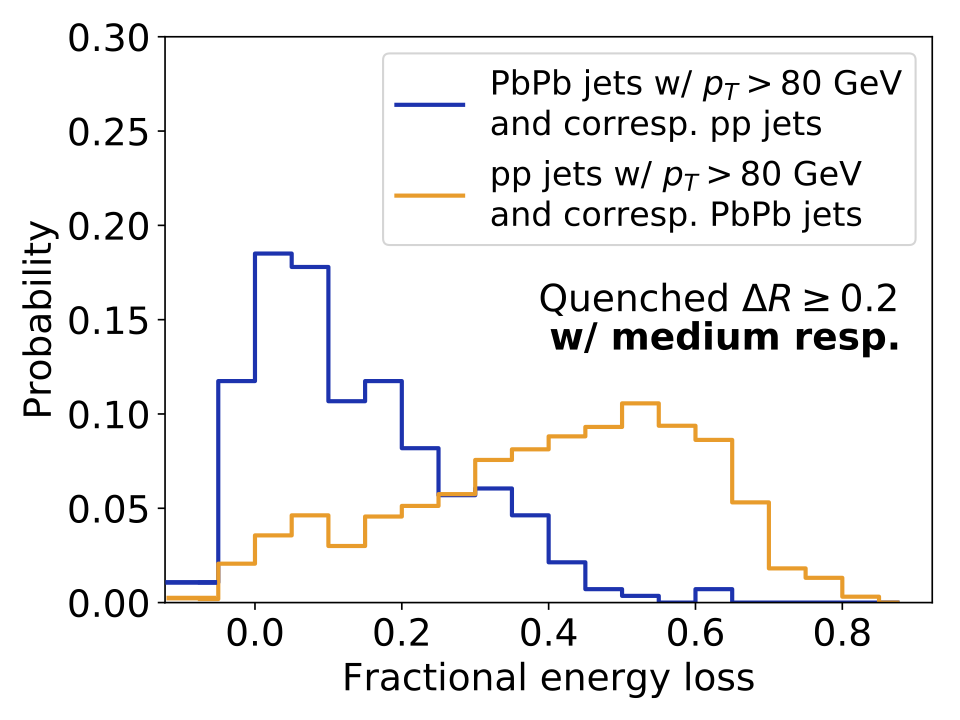}  
  \caption{}
  \label{fig:Eloss_largeR_med}
\end{subfigure}
\centering
\vspace{-0.06in}
\caption{Fractional energy loss for Quench-then-Select (blue) and Select-then-Quench (orange) Methods, without (left panels) and with (right) medium response, for $\Delta R>0.2$ (upper panels) and $\Delta R \leq 0.2$ (lower). 
}
\vspace{-0.09in}
\label{fig:Eloss}
\end{figure}

Figs.~\ref{fig:deltaR_method1_nomed} and \ref{fig:deltaR_method1_med} indeed show that the $\Delta R$ distribution {\it for jets selected in PbPb collisions} is not much modified by quenching. However, selecting heavy ion jets with $p_T>80$ GeV 
yields a sample of jets with a strong bias toward selecting those jets that do not lose much $p_T$, since the probability to produce a proton-proton jet falls rapidly with $p_T$. This bias is well known, but we confirm it in Fig.~\ref{fig:Eloss}.
What we are seeing in the upper panels of Fig.~\ref{fig:deltaR} is that the $\Delta R$ distribution of those jets which lose only a small fraction of their energy is not much modified by quenching.
In Figs.~\ref{fig:deltaR_method2_nomed} and \ref{fig:deltaR_method2_med}, we use the Select-then-Quench Method, selecting a sample of jets in proton-proton collisions and then following what happens to {\it those} jets after quenching -- regardless of how much energy they lose. By selecting unquenched jets we eliminate the selection bias favoring jets that lose little energy. And, with the selection bias eliminated, we see that quenching actually very substantially modifies the $\Delta R$ distribution -- when we include particles coming from the response of the medium to the jet.
This rather dramatic modification of the $\Delta R$ distribution is wiped out in the upper-right panel by selection bias: in the upper-right panel we are looking at jets whose energy is little modified, and so neither is their $\Delta R$.
From the lower panels of Fig.~\ref{fig:deltaR}, 
we conclude that $\Delta R$ is in fact substantially modified by quenching on a jet-by-jet basis, contrary to what one might have been tempted to conclude from the upper panels -- in which the effects of jet modification are obscured by the effects of selection bias.


In Fig.~\ref{fig:Eloss} we confirm these conclusions by looking at the distribution of the fractional energy loss of jets in the Quench-then-Select Method (blue curves) and the Select-then-Quench Method (orange curves), with and without medium response, for jets with $\Delta R<0.2$ and $\Delta R>0.2$.  In all cases, selecting quenched jets (Quench-then-Select) yields a sample of jets that have lost less energy than if one selects unquenched jets and then quenches them (Select-then-Quench). We can furthermore compare the upper (small $\Delta R$) and lower (large $\Delta R$) panels of Fig.~\ref{fig:Eloss} and see that the enhancement of jets at large $\Delta R$ caused by quenching that we see in Fig.~\ref{fig:deltaR_method2_med} is coming from 
jets that lose a substantial fraction of their energy.
These are exactly the jets that 
are missed in the Quench-then-Select Method due to its strong and intrinsic selection bias, as described earlier.


{\bf Conclusion and Discussion:} In the hybrid model, quenching modifies the Soft Drop $\Delta R$ of individual jets substantially. The jets whose $\Delta R$ is substantially modified (see Fig.~\ref{fig:deltaR_method2_med}) are those which lose a large fraction of their energy (see Fig.~\ref{fig:Eloss_largeR_med}). Selecting a jet sample using a cut on the jet $p_T$ in heavy ion collisions creates a bias towards jets that lose very little energy. These are the jets whose $\Delta R$ is not substantially modified. By selecting a jet sample using a cut on the jet $p_T$  in proton-proton collisions and considering the quenched versions of these jets, we remove the bias toward less modified jets and see that the $\Delta R$ of individual jets is substantially modified in the hybrid model. 
We furthermore see that this modification is to a significant degree coming from the response of the medium to the jet. This indicates that, unlike for jets in vacuum, the $\Delta R$ of jets in heavy ion collisions does not give us access in any straightforward way to the first hard splitting during the formation of the jet.



Because the effects of selection bias and jet modification on the observables that we have looked at are so large, we expect our qualitative conclusions to generalize far beyond the model that we have used as a testbed in which to disentangle these effects.  That said, what we have done is intrinsically a Monte Carlo study, not something that can be done in the same way in an analysis of experimental data.  Our results motivate experimental analyses of jet samples selected in a way that includes no bias toward small fractional energy loss --- for example Z+jet samples with selection cuts imposed only on the Z --- and which includes those jets that have lost a substantial fraction of their energy. It would be very interesting to look at  the $\Delta R$ distribution in such a sample.

{\bf Acknowledgments:} We are grateful to Dani Pablos for providing us with hybrid model samples and much helpful advice and to Yen-Jie Lee, Andrew Lin, Guilherme Milhano, James Mulligan, and Jesse Thaler for helpful conversations.
This work was supported by the U.S. Department of Energy, Office of Science, Office of Nuclear Physics grant DE-SC0011090.


\end{document}